\documentclass[12pt,preprint]{aastex}
\usepackage{natbib}
\usepackage{amsmath,amsfonts,amssymb}
\usepackage{graphicx}
\def\idm#1{{\mbox{\scriptsize #1}}}

\newcommand\Chi{{(\chi^2_\nu)^{1/2}}}

\def\url#1{\texttt{#1}}

\shorttitle{Substellar-mass companions to three stars}
\shortauthors{Niedzielski et al.}
\begin{document}

\title{Substellar-mass companions to the K-dwarf BD +14 4559 and the K-giants HD 240210 and BD +20 2457}

\author{A. Niedzielski\altaffilmark{1},   G. Nowak\altaffilmark{1}, M. Adam\'ow\altaffilmark{1}, A. Wolszczan\altaffilmark{2,3}}

\altaffiltext{1}{Toru\'n Center for Astronomy, Nicolaus Copernicus University, ul. Gagarina  11, 87-100 Toru\'n, Poland, Andrzej.Niedzielski@astri.uni.torun.pl,  Grzegorz.Nowak@astri.uni.torun.pl, Monika.Adamow@astri.uni.torun.pl}

\altaffiltext{2}{Department of Astronomy and Astrophysics, the Pennsylvania State University, 525 Davey Laboratory, University Park, PA 16802, alex@astro.psu.edu}

\altaffiltext{3}{Center for Exoplanets and Habitable Worlds, the Pennsylvania State University, 525 Davey Laboratory, University Park, PA 16802}

\begin{abstract}
We present the discovery of substellar-mass companions to three stars by the ongoing Penn State - Toru\' n Planet Search (PTPS) conducted with the 9.2-m Hobby-Eberly Telescope. The K2-dwarf, BD +14 4559, has a 1.5 M$_{J}$ companion with the orbital period of  269 days and shows a non-linear, long-term radial velocity trend, which indicates a possible presence of another planet-mass body in the system. The K3-giant, HD 240210, exhibits radial velocity variations that
require modeling with multiple orbits, but the available data are not yet sufficient to do it unambiguously. A tentative, one-planet model calls for a 6.9 M$_J$ planet in a 502-day orbit around the star. The most massive of the three stars, the K2-giant, BD +20 2457, whose estimated mass is 2.8$\pm$1.5 M$_\odot$, has two companions with the respective minimum masses of 21.4 M$_J$ and 12.5 M$_J$ and orbital periods of 380 and 622 days. Depending on the unknown inclinations of the orbits, the currently very uncertain mass of the star, and the dynamical properties of the system, it may represent the first detection of  two brown dwarf-mass companions orbiting a giant. The existence of such objects  will have consequences for the interpretation of the so-called brown dwarf desert known to exist in the case of solar-mass stars.

\end{abstract}


\section{Introduction}

Searches for planets around giant stars offer an efficient way to extend studies of planetary system formation and evolution to stellar masses substantially larger than 1 M$_{\odot}$ \citep{2003ApJ...597L.157S, 2006A&A...457..335H, 2007ApJ...669.1354N, 2008A&A...480..215H, 2008PASJ...60..539S, 2008PASJ...60.1317S, 2009A&A...499..935D, 2009ApJ...693..276N, 2009RAA.....9....1L}. Although searches for the most massive substellar companions to early-type stars are possible \citep{2005A&A...443..337G, 2009A&A...495..335L}, it is much more efficient to utilize the power of the radial velocity (RV) method by exploiting the fact the GK-giants, the descendants of the main sequence A-F type stars, have cool atmospheres and sufficiently many narrow spectral lines to make achieving a $<$10 m s$^{-1}$ RV measurement precision possible. 
  
The GK-giant surveys
are beginning to provide the statistics, which are needed to constrain
the efficiency of planet formation as a function of stellar mass and chemical composition. In fact, the initial analyses
by \cite{2007ApJ...665..785J} and \cite{2007A&A...472..657L} extend to giants the correlation between planetary masses and the masses of their primaries observed for the lower-mass stars. Most likely, this is simply because massive stars tend to have more massive disks. These results are in accord with
the core accretion scenario of planet formation \citep{2008ApJ...673..502K}. Furthermore, \cite{2008arXiv0801.3336P} 
have used the apparent lack of correlation between the frequency of planets
around giants and stellar metallicity to argue that this effect may imply
a pollution origin of the observed planet frequency - metallicity correlation
for main sequence stars \citep{2005ApJ...622.1102F}. Similar conclusions have been reached in a preliminary study of our giant star sample by \citep{2009arXiv0904.0374Z}.

The steadily extending baselines of the ongoing surveys of post-MS giants begin furnishing multiplanet system detections that are needed to study the dynamical evolution of planets around the off main-sequence stars. The multiplicity of planetary systems around MS stars has been firmly established \citep{2007ARA&A..45..397U, 2009ApJ...693.1084W} and there is no reason why this tendency should not be present among giants. In fact, the ongoing search for planets around GK-giants with the Hobby-Eberly Telescope (HET) by our group (Penn State-Toru\' n Planet Search, hereafter PTPS) has already tentatively identified a two-planet system around the K0-giant HD 102272 \citep{2009ApJ...693..276N}, and long-term trends in the RV data have been seen in other surveys (e.g.  \cite{2008PASJ...60..539S}).

In this paper, we describe the detection of a planet around the K3-giant, HD 240210, and show evidence that the star has more low-mass companions, and the intriguing discovery of two brown dwarf-mass bodies orbiting the K2-giant, BD +20 2457. We also describe the discovery of a Jupiter-mass planet and a non-linear RV trend in the K2-dwarf, BD +14 4559. This detection is the first result of the extension of the PTPS to evolved dwarfs in the upper envelope of the MS. This part of the project will be described in detail in a future paper.
 
The plan of this paper is as follows. An outline of the observing procedure and a description of the basic properties of the three stars are given in Section 2, followed by the analysis of radial velocity measurements in Section 3. The accompanying analysis of rotation and stellar activity indicators is given in Section 4. Finally, our results are summarized  and further discussed in Section 5.

\section{Observations and properties of the stars}

Observations were made  with the Hobby-Eberly Telescope (HET) \citep{lwr98} equipped 
with the High Resolution Spectrograph (HRS) \citep{tull98} in the queue scheduled mode \citep{HetQ}. The spectrograph was used in the R=60,000 resolution  mode with a gas cell ($I_2$) inserted into the optical path, and it was fed with a 2 arcsec fiber. 
Details of our survey, the observing procedure, and data analysis have been described in detail elsewhere \citep{2007ApJ...669.1354N, 2008IAUS..249...43N}

Radial velocities were measured 
using the standard I$_2$ cell calibration technique \citep{Butler+1996}. A template spectrum was 
constructed from a high-resolution Fourier Transform Spectrometer (FTS) I$_2$ spectrum and 
a high signal-to-noise stellar spectrum measured without the I$_2$ cell. Doppler shifts were 
derived from least-squares fits of template spectra to stellar spectra with the imprinted I$_2$ 
absorption lines.  The radial velocity for each epoch was derived as a 
mean value of the independent measurements from the 17 usable echelle orders with
a typical, intrinsic uncertainty of  6-8 m s$^{-1}$  at 1$\sigma$-level. This RV precision levels made it quite sufficient to 
use the \cite{1980A&AS...41....1S} algorithm to refer the measured RVs to the Solar System barycenter.

The long-term precision of our RV measurements has been verified by the analysis of data derived from the monitoring of stars that do not exhibit detectable RV variations. The results for the K0 giant BD+70 1068 ( V=9.29, B-V=1.15 $\pm$ 0.04, T$_{eff}$= 4580 $\pm$ 17 K) are shown in Figure 1 as an example. The relative RVs for this star vary with $\sigma$=12 m s$^{-1}$, whereas our estimated intrinsic RV uncertainty for this series of observations was 7 m s$^{-1}$. The excess RV scatter is consistent with the approximate 10 m s$^{-1}$ amplitude of solar-type oscillations derived from the scaling relations of \cite{1995A&A...293...87K}.

The atmospheric parameters of the stars under consideration were taken from Zielinski et al. (2009, in preparation, hereafter Z09), who have estimated their values using the method of \citet{2005PASJ...57...27T, 2005PASJ...57..109T}. With these values and the luminosities estimated from available data, stellar masses were derived by fitting the ensemble of parameters characterizing the star (log(L/L$_\odot$), log T$_{eff}$, log (g), and metallicity) to the evolutionary tracks of \cite{girardi2000}. The parameters of the three stars are summarized in Table 1.

\subsection {BD +14 4559}

Radial velocities of BD +14 4559  were measured at 43 epochs over the period of  1265 days between MJD 53546 and 54811.
Typically, the signal-to-noise ratio per resolution element in the spectra was 150-260 at 594 nm in 
10-25  minutes of integration, depending on the atmospheric conditions.  The estimated mean RV uncertainty for this star was 8 m s$^{-1}$

BD +14 4559 (AG +14 2370, HIP 104780, LTT 16221) is a high proper motion K5 (\cite{HIC}) star with  V=9${^m}$.63 \citep{1986AJ.....91..626W},  B-V=0${^m}$.98$\pm$0.08 \citep{2000A&A...355L..27H}   and  $\pi$= 19.99$\pm$1.61 mas \citep{1997ESASP1200.....P}. The trigonometric parallax is consistent with the photometric value of $\pi$= 24 mas derived by \citep{1986AJ.....91..626W}. The atmospheric parameters of BD +14 4559  as determined in Z09  indicate  that it is a dwarf with  log(g)=4.60$\pm$0.05, T${_{eff}}$=5008$\pm$20 K,  and [Fe/H]=0.10$\pm$0.07.  

Using the existing Hipparcos \citep{1997ESASP1200.....P} and 2MASS \citep{2006AJ....131.1163S} photometry and the empirical calibrations of \citep{2005ApJ...626..446R}, we have estimated the effective temperature of the star to be  T${_{eff}}$=4814$\pm$26 K. As this value of T$_{eff}$ is too high for a K5 dwarf, we conclude that the spectral type of BD +14 4559 is K2V.

The derived absolute bolometric magnitude of the star, M$_{V}$, and log(L/L$_\odot$), are 5.56 and -0.32, respectively. By comparing the stellar parameters to evolutionary tracks of \cite{girardi2000}, we have computed the mass of BD +14 4559 to be M/M$_\odot$=0.86 $\pm$ 0.15  and R/R$\odot$= 0.95 $\pm$ 0.2.

The projected rotational velocity of BD +14 4559 , $v sin i= $2.5 $\pm$ 1 km s$^{-1}$, was estimated using the cross-correlation method \citep{Benz+Mayor1984}. From this value and the adopted stellar radius we have obtained an estimate of the rotation period of P$_{rot}$=19 days, which is much shorter than the observed 267-day period of the RV variations. From the uncertainty of the $v sin(i)$ determination alone, the stellar rotation period may range from 13 to 32 days.

\subsection{HD 240210}

Radial velocities of HD 240210  were measured at 38 epochs over the period of  1655 days between MJD 53187  and 54842.
Typically, the signal-to-noise ratio per resolution element in the spectra was 150-300 at 594 nm in
4-10  minutes of integration, depending on the atmospheric conditions.  The estimated mean RV uncertainty for this star was 8 m s$^{-1}$

HD 240210  (BD +56 2959) is a  K7 \citep{hd} star with  V=8${^m}$.33, B-V=1${^m}$.The parallax of 6.5$\pm$0.04 \citep{2000A&A...355L..27H}   and $\pi$= 7.6$\pm$8.7 mas \citep{1997ESASP1200.....P} are consistent with the $\pi$= 7.0$\pm$2.6 mas  of \citep{1992AJ....104.1237G}.
The atmospheric parameters of HD 240210  (Z09)   indicate that the star is a giant with  log(g)=2.31$\pm$0.11, T${_{eff}}$=4297$\pm$25 K, and [Fe/H]=-0.18$\pm$0.12.

Using the existing Hipparcos \citep{1997ESASP1200.....P} and 2MASS \citep{2006AJ....131.1163S} photometry, and the empirical calibrations of \citep{2005ApJ...626..446R}, we have estimated the effective temperature of the star to be  T${_{eff}}$=4290$\pm$13 K. These values of effective temperature and log(g) are consistent with a K3-giant star.

The derived absolute bolometric magnitude of the star and log(L/L$_\odot$), are  0.38 and 1.75, respectively. By comparing stellar parameters to evolutionary tracks of \cite{girardi2000} we have computed the mass of HD 240210 to be M/M$_\odot$=1.25 $\pm$ 0.25 and R/R$\odot$= 13 $\pm$ 3.

The projected rotational velocity of HD 240210 , $v sin i<$ 1.0 $\pm$ 1 km s$^{-1}$, was estimated using the cross-correlation method. From this value and the adopted stellar radius we have obtained an estimate of the rotation period of P$_{rot}>$ 654    days.

\subsection{BD +20 2457}

Radial velocities of BD +20 2457  were measured at 37  epochs over the period of  1833 days between MJD 53033 and 54866.
Typically, the signal-to-noise ratio per resolution element in the spectra was 140-260 at 594 nm in
12-30  minutes of integration, depending on the atmospheric conditions.  The estimated mean RV uncertainty for this star was 7 m s$^{-1}$

BD +20 2457  (AG +20 1166) is a K2 \citep{1975ascp.book.....H} star
with  V=9${^m}$.75, B-V=1${^m}$.25$\pm$0.09 \citep{2000A&A...355L..27H} which is similar to V=9.57 and B-V=1.35 as measured photometrically by \citep{1980IBVS.1898....1B}, who observed it as a comparison star in their stellar variability study. The parallax of the star, $\pi$= 5.0$\pm$26.50 mas \citep{1997ESASP1200.....P}, is obviously very poorly determined.  The atmospheric parameters of BD +14 4559 (Z09) indicate  that it is a dwarf with log(g)=1.51$\pm$0.05, T${_{eff}}$=4137$\pm$10 K,  and [Fe/H]=-1.00$\pm$0.07.

Using the existing Hipparcos \citep{1997ESASP1200.....P} and 2MASS \citep{2006AJ....131.1163S} photometry and the empirical calibrations of \citep{2005ApJ...626..446R}, we have estimated the effective temperature of the star to be  T${_{eff}}$=4127$\pm$17 K. We conclude that the spectral type of BD +20 2457 is K2II.

The derived absolute bolometric magnitude of the star, M$_{V}$, and log(L/L$_\odot$), are -3.0 and 3.17, respectively. By comparing the stellar parameters to evolutionary tracks of \cite{girardi2000}, we have estimated the mass of BD +20 2457 to fall in the 1.3 - 4.3 M/M$_\odot$ range (2.8$\pm$1.5 M/M$_\odot$),   and R/R$\odot$= 49 $\pm$ 26.

As for the other two stars, the projected rotational velocity of BD +20 2457 , $v sin i< $1.0  km s$^{-1}$, was estimated using the cross-correlation method. From this value and the adopted stellar radius we have obtained an estimate of the rotation period of the star, P$_{rot}>$ 2460   days.

\section{Analysis of the radial velocity data}

The process of modeling the RV measurements of the three stars discussed in this paper is documented in Figures 2-6. In the first step of the analysis, single, 6-parameter, Keplerian orbits were least-squares fitted to data using the Levenberg-Marquardt algorithm \citep{1992nrfa.book.....P}. For the K-giants, HD 240210 and BD +20 2457, the estimated, 7-8 m s$^{-1}$ errors in RV measurements were clearly insufficient to account for the observed RV variations in the post-fit residuals. This excess RV variability is believed to originate in fluctuations of the stellar surface and radial and non-radial oscillations of the giant stars \citep{2009Natur.459..398D} with the typical RV amplitudes around 20 m s$^{-1}$ \citep{2006A&A...454..943H}. To account for this effect, we have quadratically added 10 m s$^{-1}$ and 30 m s$^{-1}$ to the calculated RV errors for HD 240210 and BD +20 2457, respectively.

Evidently, in all the three cases, the best-fit residuals from single orbit models leave non-random trends that need to be modeled further. A statistical significance of these trends was assessed by calculating false alarm probabilities (FAP) for a null hypothesis that the trends can be adequately accounted for by a single Keplerian orbit and noise. The FAPs were calculated with the aid of the radial velocity scrambling method (\cite{2007ApJ...657..533W}, and references therein). As illustrated in Figure 6, FAP $<$ 0.1\% for the three data sets, which indicates that the additional, systematic RV variations seen in the data are statistically significant.

\subsection{BD +14 4559}

As shown in Figure 3, the best-fit, single Keplerian orbit model for the RV variations of this star leaves behind a significant, non-linear trend, which reaches the amplitude of $\sim$50 m s$^{-1}$ over the 1300-day span of observations. At this point, rather than attempting to search the $\chi^2$- space for a range of acceptable two-planet solutions, we have modeled the observed RV variations with a Keplerian orbit and a parabolic trend added to it. The best-fit $\chi^2$ value for this model drops from 4.4 for the single orbit case to 2.1 indicating, together with the FAP $<$0.1\%, that the trend is significant.

The final solution (Table 1) includes a planet with the minimum mass of 1.5 M$_{Jup}$, in  a 269-day, 0.78 AU, $e=$0.29 orbit around the star. Given the stellar mass of 0.86 M$_\odot$, the mass function of 2.5 $\times$ 10$^{-8}$ M$_\odot$, and the above RV amplitude and time span of the observed trend, one can broadly constrain a putative second companion to be another planet with the minimum mass $>$2.4 M$_{Jup}$ and the orbital radius $>$2.3 AU, assuming $e=$0.

The remaining post-fit rms residual, $\sigma_{RV}$=11 m s$^{-1}$, slightly exceeds the estimated 8 m s$^{-1}$ precision of our RV measurements for this star. This excess "jitter" is within the estimated, intrinsically generated RV noise of 5 m s$^{-1}$ due to solar-type oscillations in stable dwarfs \citep{2005PASP..117..657W}. 

\subsection{HD 240210}

A single orbit fit to the RV data for this star (Figure 4, Table 2) gives $\chi^2$=9.8, the post-fit rms residual, $\sigma_{RV}$=39m s$^{-1}$, and the very clear unmodeled RV variations. Our search for a multiple orbit solution has resulted in several two- and three-planet models, which gave nearly identical improvements of the fit. This is illustrated in Figure 4 by a two-planet model, which produces $\chi^2$=5.0, $\sigma_{RV}$=25 m s$^{-1}$, FAP $<$ 0.1\%, and the persistent lower-level systematics in the RV residuals. Because the existing measurements are obviously insufficient to obtain satisfactory constraints on the Keplerian model of the RV variations of this star, we list in Table 2 the initial one-planet solution and postpone further analysis until the time, when enough data become avaiable. The provisional, best-fit orbit involves a 6.9 M$_{Jup}$ planet in a 502-day orbit with the semi-major axis of 1.3 AU and $e=$0.15.
 
We also note that the 25 m s$^{-1}$ post-fit residual from the multiplanet fits is about 3 times larger than the formal precision of the RV measurements for HD 240210. As mentioned above, this is in agreement with the estimated RV scatter in K-giants due to their internal activity.

\subsection{BD +20 2457} 

A single, Keplerian orbit fit to the RV data for this star is shown in Figure 4. The residuals from this fit are characterized by $\chi^2$=13.7, $\sigma_{RV}$=105 m s$^{-1}$, and a $\sim$600-day periodicity with the amplitude in excess of 100 m s$^{1-}$. Once again, as shown in Figure 6, the FAP for the existence of the second periodicity is less than 0.1\%.

As shown in Figure 5, the best-fit, two-orbit solution offers a dramatic improvement to the Keplerian model of the RV variations observed in BD +20 2457, with the values of $\chi^2$ and $\sigma_{RV}$ reduced to 5.2 and 60 m s$^{-1}$, respectively. The model parameters calculated for the stellar mass of 2.8 M$_\odot$ and listed in Table 3 call for a system of two substellar-mass companions with the respective minimum masses of 21.4 M$_{Jup}$ and 12.5 M$_{Jup}$, orbital periods of 380 and 622 days, semi-major axes of 1.45 and 2.01 AU, and mild eccentricities of 0.15 and 0.18. 

As in the case of HD 240210, the estimated, 7 m s$^{-1}$ uncertainty of the RV measurements of BD +20 2457 is much smaller than the actual, 60 m s$^{-1}$ rms of the post-fit residuals. This additional RV "jitter" broadly agrees with the 138 m s$^{-1}$ amplitude of the solar-type oscillations extrapolated for this star from \cite{1995A&A...293...87K}.

\section{Stellar photometry, rotation, and line bisector analysis}

In order to verify that the observed RV periodicities are indeed caused by the Keplerian motion, we have thoroughly examined the existing photometry data in search for any periodic light variations, and performed a complete analysis of line bisectors and curvatures for each of the three stars discussed in this paper. This analysis used the cross-correlation method proposed by \citep {MartinezFiorenzano2005}. For each star, the cross-correlation functions were computed from $\sim$1000 line profiles with the I$_2$ lines removed from the spectra. This procedure has been described in detail in \cite{Nowak-torun}.

Because both the Ca II K emission line and the infrared Ca II triplet lines at 849.8-854.2 nm are outside the range of our spectra, we have used the H$\alpha$ line as a chromospheric activity indicator.  As the shape of the H$\alpha$ line in the spectra that showed significant telluric line contamination could be affected by it, care was taken to omit such spectra from the analysis. Equivalent widths (EW) of the line profile, defined as I/I$_{c}$$\leq$0.9, were measured within the central 1 $\AA$ of the absorption profile, where the signature  of stellar activity should be most pronounced (the uncertainty of the EW defined this way depends on the  actual depth of  the H$\alpha$ line and the signal-to-noise ratio).

\subsection{BD +14 4559}

The existing photometric databases for this star cover a wide range of epochs from, MJD 47891 to  54792, and include extensive time series of measurements from the  Hipparcos and Tycho \citep{1997ESASP1200.....P}, the NSVS \citep{2004AJ....127.2436W}, and the ASAS experiment \citep{1997AcA....47..467P}. None of the four data sets reveal any periodic light variations. In particular, the ASAS data, which are contemporaneous with our RV measurements of BD +14 4559, include observations made at 280 epochs between MJD 52754 and MJD 54792. In this case, the light variations are characterized by the mean value of V=9.650$\pm$0.021 mag, which is consistent with the earlier measurements. As an example, the Lomb-Scargle (LS) periodogram of the ASAS photometry data in shown in Figure 7.

We have obtained  43  bisector and line curvature measurements, which resulted in the mean bisector velocity span,  BVS=4.6 $\pm$ 32 m s$^{-1}$, and the mean bisector curvature, BC=-8 $\pm$ 27 m s$^{-1}$ with no trace of any periodic variability.  No correlation between the BVS, BC, and RV was found (r=0.09 and r=0.08, respectively). Similarly, our analysis of  23 spectra of the star has shown that the EW of the H$_{\alpha}$ line exhibited a 3\% rms scatter around its mean value of $396\pm 12 m\AA$, which was not correlated with the observed RV variability (r = -0.15, while the critical value of the Pearson correlation coefficient at the confidence level of 0.01 is  r$_{21, 0.01}$=0.53). The LS periodograms of  the H$_{\alpha}$  EW , the BVS, and BC for BD +14 4559  are shown in Figure 7. Clearly, the only observable that exhibits periodic time variations in time is the radial velocity.

Using the scatter seen in the ASAS photometry of the star, and its rotational velocity determined above, we can estimate the amplitude of radial velocity variations and of the BVS due to a possible presence of a spot on the stellar surface \citep{Hatzes2002}. The observed  radial velocity amplitude of BD +14 4559 is  three times larger than the 35 m s$^{-1}$ total RV amplitude predicted by \cite{Hatzes2002} for a spot with  a filling factor of f=0.02, on a star rotating at $v sin i=$2.5 km s$^{-1}$. Within the estimated uncertainty of the rotation velocity, the expected RV amplitude induced by a hypothetical spot is at least two times less than the observed one. The expected  bisector variations of  5 m s$^{-1}$ are comparable to the precision of our RV measurements and cannot have a detectable effect on our results. To summarize, the facts that there is no variability in the available photometric measurements over the period of $\sim$19 years, the observed period of the RV v
 ariations is much longer than the estimated rotation period of the star, a significant eccentricity must be included in the best-fit model of the RV curve, and that the observed RV variations have been consistent over the five consecutive cycles of the period, make their interpretation in terms of a rotating spot very unlikely.

\subsection{HD 240210}

In the case of this star, the only long-term photometry available comes from 129 Tycho measurements with the V$_{T}$ filter   
\citep{1997ESASP1200.....P}, between MJD 47946  and 49016. These data give the mean V$_{T}$=8.499 $\pm$ 0.012 and the scatter in V$_{T}$ of  0.136 mag.   No periodic variability has been detected in these measurements.  
In addition, there are 19  epochs of photometric observations of this star available from the NSVS \citep{2004AJ....127.2436W}, collected between MJD 51362 and 51493. These data are characterized by the median V=7.965 $\pm$ 0.01 and the scatter of 0.038. Since the NSVS photometry is of a much better quality than the Tycho data, we consider this scatter more believable.
As shown in Figure 8, the LS periodogram of the Tycho photometry of HD 240210 reveals no significant periodicities.

The  mean values of the BVS and the BC for this star are 62 $\pm$ 23 m s$^{-1}$, and 23 $\pm$ 27 m sÃ, respectively,  with no statistically significant variability. No correlations between the RV residuals and the BVS and BC were found. The respective correlation coefficients  are r=0.08  for the BVS and r=-0.18  for the BC  (r$_{36, 0.01}$=0.40). Furthermore, the average EW of the H$_{\alpha}$ line measured over 28 epochs amounts to 602 $\pm$ 13, and the EW itself does not show any periodicities within  2$\%$.  We have considered a possible rotational modulation in H$_{\alpha}$ by calculating the Pearson correlation coefficient with the RV data residuals after removing  the putative orbit. We found no correlation between  the H$_{\alpha}$  EW and the RV residuals (r= -0.26, whereas r$_{26, 0.01}$=0.48). The LS periodograms of the four observables for HD 240210 discussed above are shown in Figure 8. As in the case of BD +14 4559, radial velocity is the only one that varies periodical
 ly. 

The observed photometric scatter in HD 24210, if interpreted in terms of a spot rotating with the star would also be detectable in both the radial velocity and the line bisector data. The estimated total RV variations should be of the order of 23 m s$^{-1}$ and the bisector variations should amount to 31 m s$^{-1}$ \citep{Hatzes2002}.  The semi-amplitude of the observed periodic signal in radial velocity is much larger than the predicted variations induced by a hypothetical spot of size consistent with the existing photometeric  variations.  Periodic variations in line bisectors  produced by a hypothetical spot  should manifest themselves as the uncertainty in our  BVS measurements, which is comparable the the predicted value. In any case, these variations should show a periodicity equal to the stellar rotation period, P $>$ 654 d. Because no such variations are present in our bisector measurements for this star, we conclude that the observed scatter is a combination of low quality photome
 try and possible solar type oscillations of the star. Such oscillations might account for 3.3 mmag in the photometric variability and V$_{osc}$ = 13 m s$^{-1}$ in the BVS using the scaling relations of \cite{1995A&A...293...87K}.

\subsection{BD +20 2457}

As in the case of BD +14 4559, we have identified four photometric data sets that we could utilize in our analysis, but we have concentrated on the data from the ASAS project \citep{1997AcA....47..467P}, which contains extensive measurements that are contemporaneous with our RV observations. These data include 120 measurements made between MJD 52622 and MJD 53400 and the 191 additional ones made between MJD 52623 and MJD 54911. We have used 319 grade A measurements from the ASAS to derive the mean V=9.693 $\pm$ 0.051 mag and to compute their LS periodogram in search for any periodicities. As shown in Figure 9, no significant periodicities are present in the spectrum. Similarly, no periodicities were found in the other three, lower quality time series.

We have obtained 37  bisector and curvature measurements for this star, with the mean BVS=105 $\pm$ 39 m s$^{-1}$, and the mean BC=47 $\pm$ 30 m s$^{-1}$, both revealing no significant periodicities in the corresponding LS periodograms (Figure 9).  
We have also searched for  correlations between the RV residuals and the BVS and BC after having removed the putative companions b and c. The respective correlation coefficients are  r=-0.33 and r=-0.19 for BVS and r=-0.12 and r=-0.09 for BC, respectively (the critical value of Pearson correlation coefficient at the confidence level of 0.01 is r$_{35, 0.01}$=0.42). 
The EQ of the H$_{\alpha}$ line for BD +20 2457 is 653$\pm$ 23, and this parameter shows no significant periodicities within  3.5$\%$ for 23 observations (Figure 9). As for the other two stars, we have searched for a possible rotational modulation in H$_{\alpha}$ by calculating the Pearson correlations coefficient with the RV data residuals after removing the individual orbits. We found that there is no significant  correlation between  the H$_{\alpha}$  EW and the RV residuals (r=-0.21 and -0.51, respectively, and r$_{21, 0.01}$=0.53). 

A photometric scatter resulting from a spot rotating with BD +20 2457 would induce the estimated total RV variations on the order of 30 m s$^{-1}$ and the bisector variations of  40 m s$^{-1}$ \citep{Hatzes2002}.    The semi-amplitude of the observed periodic signal in radial velocities is over one order of magnitude larger.   As in HD 240210, the anticipated variability in line bisectors produced by a periodic spot rotation is similar to that observed  in our BVS measurements, these variations should show a periodicity equal to the star's rotation period, P $>$ 2640 d, which appears not to be  the case (our observations cover only  1833 days of the P $>$ 2640 day rotation period). Since no measurable variations are present in the bisectors and line curvatures, we conclude that the observed photometric scatter is again a combination of low quality photometry and possible solar type oscillations of the star. Such oscillations might account for 5.9 mmag  and up to V$_{osc}$ = 138 m s$^{-1}$ using the scaling relations of \cite{1995A&A...293...87K}.

\section{Discussion}

The most recent results from the Penn State-Toru\' n search for extrasolar planets described in this paper add a Jupiter-mass planet to the large body of planets around main-sequence stars \citep{2008NewAR..52..154S} and increase the still modest number of substellar-mass objects around GK-giants (e.g.  \cite{2009A&A...499..935D,   2009RAA.....9....1L, 2009ApJ...693..276N, 2008PASJ...60.1317S,  2007ApJ...669.1354N} and references therein) by at least three new detections.

In principle, the long period RV variations in red giants may be related to non-Keplerian effects, including the
stellar rotation modulation of surface inhomogeneities, intrinsic activity, or non-radial pulsations \citep[e.g.][]{2006A&A...457..335H}. Our analysis of the available photometric data
and the behavior of line bisectors in the three stars, following the established practices \citep{Quelozetal2001}, has shown no significant correlations between the time variability of these stellar activity indicators and the RV variations.  Consequently, our RV measurements find the most plausible explanation in terms of the Keplerian motion of substellar-mass companions around the observed stars. The details of this analysis
can be found in \cite{2008IAUS..249...49N}.

In the case of the K2-dwarf, BD +14 4559, the RV data reveal a planet with the minimum mass of 1.5 M${_J}$ in a 269-day, 0.78 AU, e=0.29 orbit around the star. An additional, long-term trend seen in the data suggests the presence of another, long-period companion, which is very likely to be another planet. This adds to the growing number of confirmed and anticipated multiplanet systems around solar-type stars \citep{2009ApJ...693.1084W}. 

A single planet model of the RV variations in the K3-giant, HD 24210, leaves highly correlated post-fit residuals signaling a possible presence of additional substellar-mass bodies orbiting the star. The presently available data are not sufficient to obtain an unambiguous multibody solution for this system. The provisional parameters for one planet that can be fitted for give a 6.9 M$_{J}$ body in a 502-day, 1.33 AU, e=0.15 orbit that will have to be revised, when another planet (or planets) are added to the current model. Nevertheless, the solution is robust enough to conclude that HD 24210 has at least one planet with large minimum mass. This result further emphasizes the observed correlation between the stellar mass and the masses of orbiting planets \citep{2007ApJ...665..785J, 2007A&A...472..657L}.

The RV variations observed in the K2-giant, BD +20 2457, can be unambiguously modeled by two Keplerian orbits with the periods of 380 and 622 days, 1.4 and 2 AU semi-major axes, and the respective eccentricities of 0.15 and 0.18. For the estimated stellar mass of 2.8 M$_\odot$, these parameters yield minimum masses of the orbiting bodies of 12.5 M$_{J}$ and 21.4 M$_{J}$, respectively. The $\sim$ 1.6 ratio of the orbital periods places the orbits relatively close to the 3:2 mean motion resonance (MMR). These characteristics suggest that the two bodies share a common origin from a disk that once surrounded the primary star. At the same time, at least one of its companions, and very possibly both of them, have masses that exceed the 13 M$_{J}$ deuterium burning limit that conventionally separates planets from brown dwarfs.

Further analysis of this intriguing system is complicated by the fact that, due to the highly uncertain parallax of the primary, its  mass is poorly known and spans a range that is at least as wide as 1.3-4.3 M$_{\odot}$ (Section 2). This is illustrated in Figure 10, which shows the constraints on the masses of the two substellar companions under the usual assumption of a random distribution of orbital inclinations. There is a 95\% probability that the masses of the two bodies fall in the respective ranges delimited by the inclinations of 18$^\circ$ and 90$^\circ$. Clearly, high masses, approaching the hydrogen burning limit are very unlikely. Similarly, a possibility that both masses stay below the deuterium burning limit is marginal at best, unless one accepts an unrealistically low primary mass, approaching 1 M$_{\odot}$ or even less. Consequently, within the currently available constraints, it is reasonable to assume that the true masses of the two companions to BD +20 24
 57 have masses in the brown dwarf range. An extensive study of the dynamics of this system will be described in a subsequent paper.

The two brown dwarf - mass bodies discovered by our survey around BD +20 2457 add to the previous three detections of single objects of this kind around intermediate-mass stars \citep{2005A&A...437..743H, 2007A&A...472..657L, 2008ApJ...672..553L}. Only one such two-companion system has been previously identified around a solar-mass star, HD 168443, by \cite{2001ApJ...555..418M}. On the basis of the facts presented above, it appears entirely reasonable to assume that the two companions to this star have originated from a massive disk and acquired enough mass to place them above the formal 13 M$_{J}$ brown dwarf limit. This raises an intriguing possibility that, in the case of substellar-mass companions to giants, both the concept of the brown dwarf desert \citep{2000PASP..112..137M} and the meaning of the deuterium burning limit will have to be revisited. In fact, this problem has been remarked on by \cite{2007A&A...472..657L}in the context of the observed correlation between the stellar and the planetary mass.

\acknowledgments
We thank  the HET resident astronomers and telescope operators for support. AN  and GN were 
supported in part by the Polish Ministry of Science and Higher Education grant 1P03D 007 30.
AW acknowledges support from NASA grant NNX09AB36G. 
GN is the recipient of a graduate stipend of the Chairman of the Polish Academy of Sciences.
The Hobby-Eberly Telescope (HET) is a joint project of the University of Texas at Austin, the Pennsylvania State University, Stanford University, Ludwig-Maximilians-Universit\"at M\"unchen, and Georg-August-Universit\"at G\"ottingen. The HET is named in honor of its principal benefactors, William P. Hobby and Robert E. Eberly.

%
% REFERENCES
%

\clearpage

%
% TABLES
%

\begin{deluxetable}{llll}
\label{tab:table1}
%\tabletypesize{\scriptsize}
%\rotate
\tablecaption{Stellar parameters of BD +14 4559, HD 24210 and BD +20 2457.}
\tablewidth{0pt}
\tablehead{ \colhead{Parameter} & BD +14 4559 & HD 24210 & BD +20 2457\\}
\startdata
V 					& 9.63 				& 8.33			& 9.75	\\
B-V 					& 0.98$\pm$0.08		& 1.65$\pm$0.04	& 1.25$\pm$0.09	\\
Spectral type 			& K2V				& K3III			&  K2II	\\
$\pi$ [mas] 			& 19.99$\pm$1.61		& 7.0$\pm$2.6		& 5.0$\pm$26	\\
T${_{eff}}$ [K]			&  5008$\pm$20		& 4290$\pm$13	& 4127$\pm$17\\
log g 				& 4.60$\pm$0.05 		& 2.31$\pm$0.11	& 1.51$\pm$0.05\\
$[$Fe$/$H$]$ 			& 0.10$\pm$0.07		& -0.18$\pm$0.12	& -1.00$\pm$0.07\\
log L$/$L$_{\odot}$  	& -0.32				& 1.75			& 3.17\\
M$_{\star}/$M${_\odot}$ & 0.86$\pm$0.15		& 1.25$\pm$0.25	& 1.3-4.3 \\
\enddata
\end{deluxetable}
\begin{table}
\label{tab:table2}
\caption{
Orbital parameters of the BD +14 4559  planet.
}
\centering
\begin{tabular}{lll}
\hline
\hline
Parameter \hspace{1em}  
& BD~+14 4559 ~{\bf b}   
\\
\hline
P~[days]                &    268.94    $\pm$  0.99  
\\
$T_0$~[MJD]             & 53293.71  $\pm$ 6.35
\\
K~[m\,s$^{-1}$]         &   55.21  $\pm$  2.29
\\
e                       &     0.29 $\pm$  0.03
\\
$\omega$ [deg]          &   87.64  $\pm$  7.87
\\
$m_2\sin~i$ [M$_{\idm{J}}$]     
                        &     1.47 
\\
$a$ [AU] 		&     0.777
\\
$\Chi$  		& {2.10}
\\
$\sigma_{RV}$~[m\,s$^{-1}$] 	& {11.43}
\\
\hline
\end{tabular}
\end{table}

\begin{table}
\label{tab:table3}
\caption{
Tentative orbital parameters of the new planet around HD 240210. 
}
\centering
\begin{tabular}{ll}
\hline
\hline
Parameter \hspace{1em}  
& HD~240210~{\bf b}   \\
\hline
P~[days]                &    501.75    $\pm$  2.33  \\
$T_0$~[MJD]             & 54486.76 $\pm$ 10.93  \\
K~[m\,s$^{-1}$]         &   161.89 $\pm$  3.49   \\
e                       &     0.15$\pm$  0.02\\
$\omega$ [deg]          &   277.49  $\pm$  7.77  \\
$m_2\sin~i$ [M$_{\idm{J}}$]     &    6.90  \\
$a$ [AU] 		&     1.33   \\
$\Chi$  		& 9.79 \\
$\sigma_{RV}$~[m\,s$^{-1}$] 	& 38.9  \\
\hline
\end{tabular}
\end{table}

\begin{table}
\label{tab:table4}
\caption{
Orbital parameters of the sub-stellar-mass companions to BD +20 2457
derived from the best fit of a two-body Keplerian model to the RV data.
}
\centering
\begin{tabular}{lll}
\hline
\hline
Parameter \hspace{1em}  
& BD +20 2457~{\bf b}  & BD +20 2457~{\bf c}   
\\
\hline
P~[days]                &    379.63    $\pm$  2.01  &  621.99    $\pm$ 10.20 
\\
$T_0$~[MJD]             & 54677.03 $\pm$ 28.19 & 53866.95 $\pm$ 27.99 \\
K~[m\,s$^{-1}$]         &   322.35 $\pm$  9.57  & 160.03   $\pm$ 7.21  
\\
e                       &     0.15 $\pm$  0.03 & 0.18  $\pm$  0.06\\
$\omega$ [deg]          &   207.64  $\pm$  21.99 & 126.02   $\pm$  16.54 \\
$m_2\sin~i$ [M$_{\idm{J}}$]     &     21.42  & 12.47
\\
$a$ [AU] 		&     1.45  & 2.01
\\
$\Chi$  		&\multicolumn{2}{c}{5.20} \\ 
$\sigma_{RV}$~[m\,s$^{-1}$] 	& \multicolumn{2}{c}{60.02}  \\
\hline
\end{tabular}
\end{table}

\clearpage

%
% FIGURES WITH CAPTIONS
%

\begin{figure}
\includegraphics[angle=0,scale=0.60]{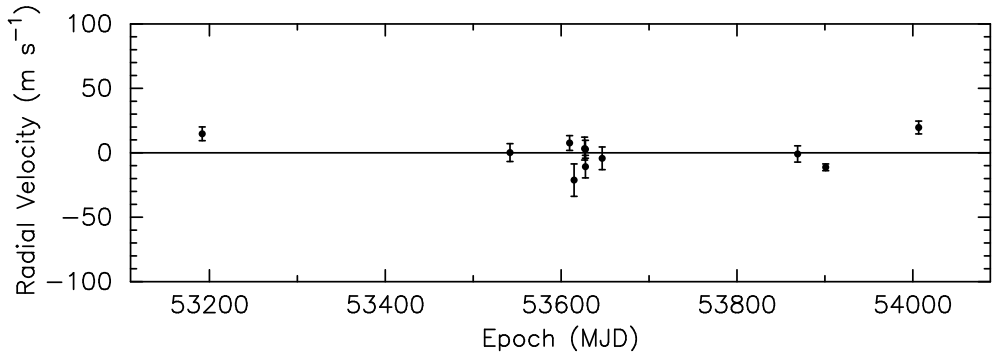}
\caption{Relative radial velocities for K0-giant BD+70 1068. The star shows  RV variability at $\sigma$=12 m s$^{-1}$ whereas the intrinsic RV measurement uncertainty for these observations was 7 m s$^{-1}$. The remaining RV scatter can be attributed to solar - type oscillations at the level of 10 m s$^{-1}$}
\end{figure}

\begin{figure}
\includegraphics[angle=-90,scale=0.60]{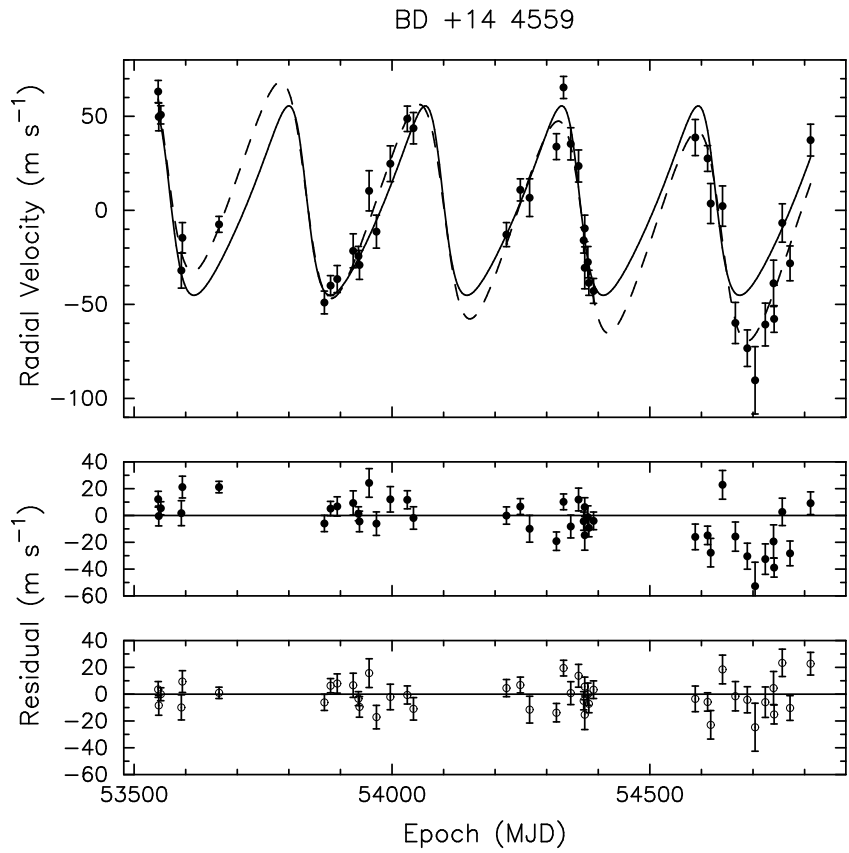}
\caption{{\sl Top:} Radial velocity measurements of BD +14 4559  (filled circles), the best-fit of a single planet Keplerian model to data (solid line) and the fit of that model with a parabolic trend (dashed line). {\sl Center:} The post-fit residuals for a single planet model. {\sl Bottom:} As above for the planet with the trend fitted out.
}
\end{figure}

\begin{figure}
\includegraphics[angle=-90,scale=0.60]{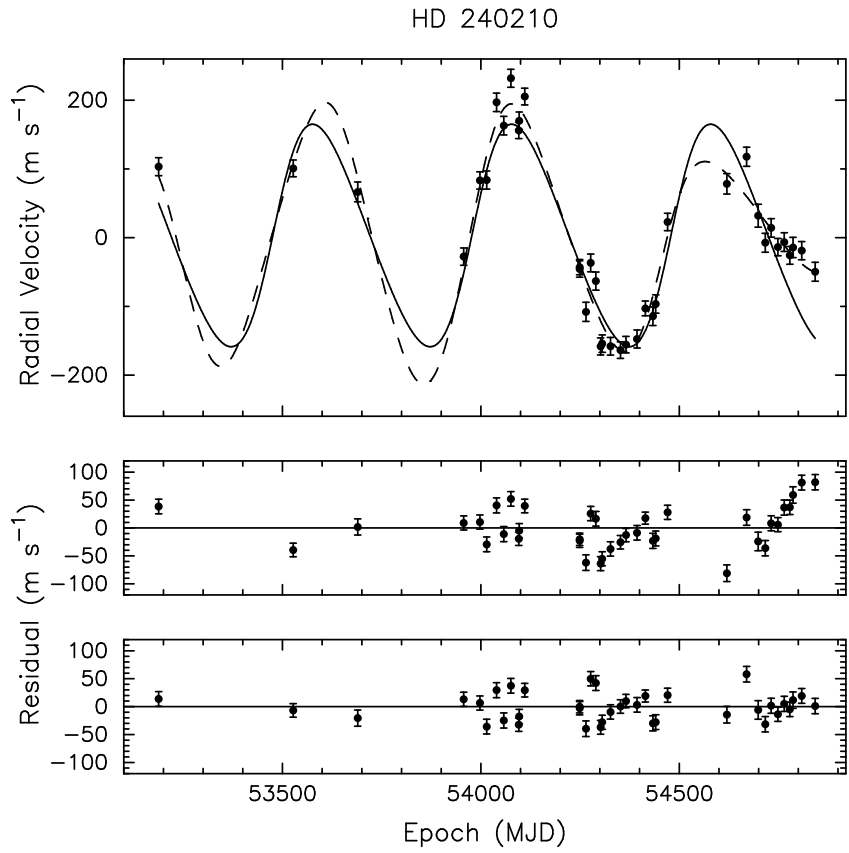}
\caption{{\sl Top:} Radial velocity measurements of HD 240210 (filled circles), the best-fit single planet Keplerian model (solid line), and a tentative, best-fit model of two planets to data (dashed line). {\sl Center and bottom:} See Figure 2.}
\end{figure}

\begin{figure}
\includegraphics[angle=-90,scale=0.60]{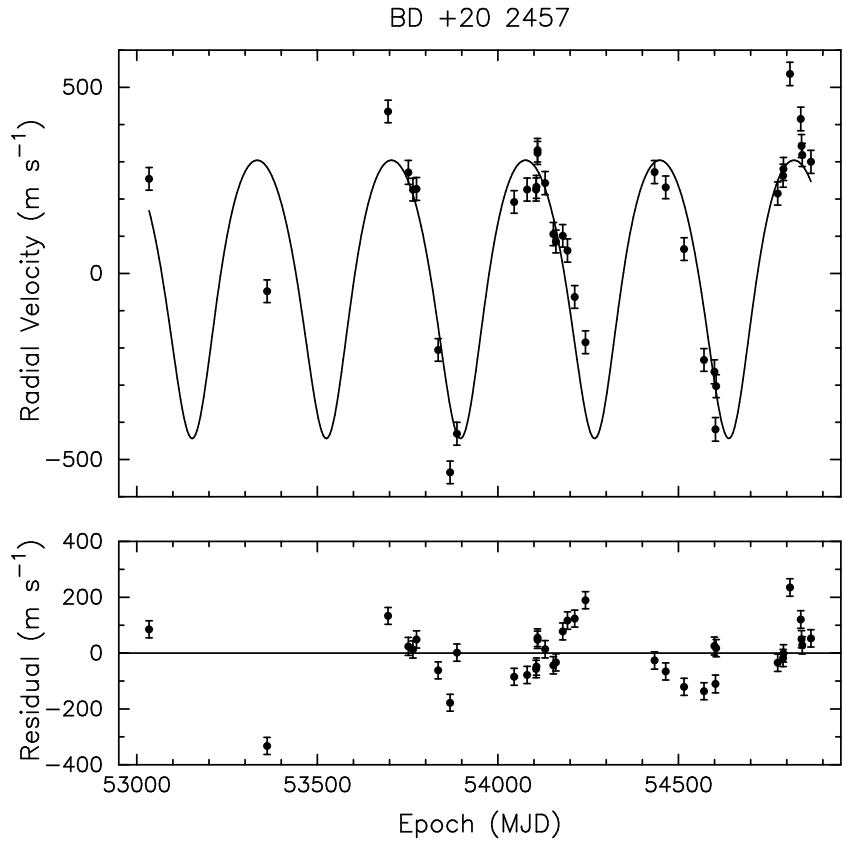}
\caption{{\sl Top:} Radial velocity measurements of BD +20 2457  (filled circles) and the best-fit of a single planet Keplerian model to data (solid line). {\sl Bottom:} The post-fit residuals for the above model.}
\end{figure}

\begin{figure}
\includegraphics[angle=-90,scale=0.60]{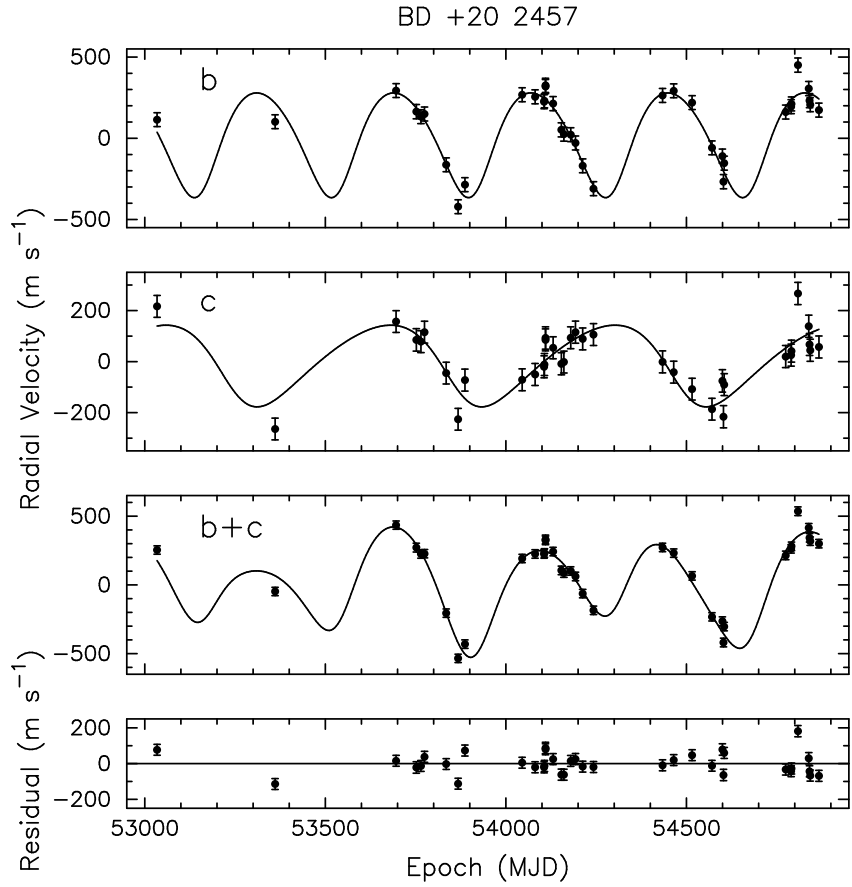}
\caption{A decomposition of the observed RV variations in BD +20 2457 (filled circles) into contributions from the orbital motions of two brown dwarf-mass companions. {\sl From top to bottom:} The observed radial velocities (filled circles) and the best-fit Keplerian orbit for companion {\sl b} (solid line) with the contribution from companion {\sl c} fitted out. The same for companion {\sl c} with companion {\sl b} fitted out. The best fit of both {\sl b} and {\sl c} to the RV data. The post-fit residuals from the two-companion fit.}
\end{figure}

\begin{figure}
\includegraphics[angle=-90,scale=0.60]{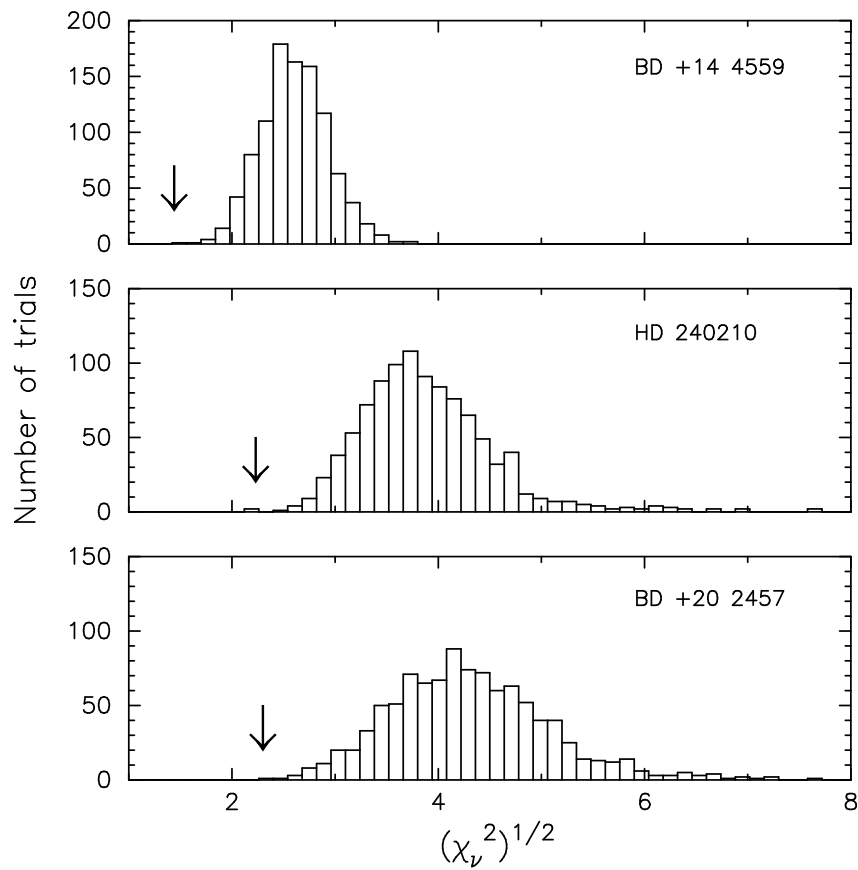}
\caption{Histograms of the values of $\sqrt{\chi^2_\nu}$ obtained from the fits of the Keplerian models to scrambled RVs used to estimate the FAPs for the three stars discussed in the text. In each case, 1000 sets of scrambled RVs have been generated. Vertical arrows point to the respective $\sqrt{\chi^2_\nu}$ values derived from fits to unscrambled velocities. In all three cases the FAP values are less than 0.1\%.}
\end{figure}

\begin{figure}
\includegraphics[angle=0,scale=0.60]{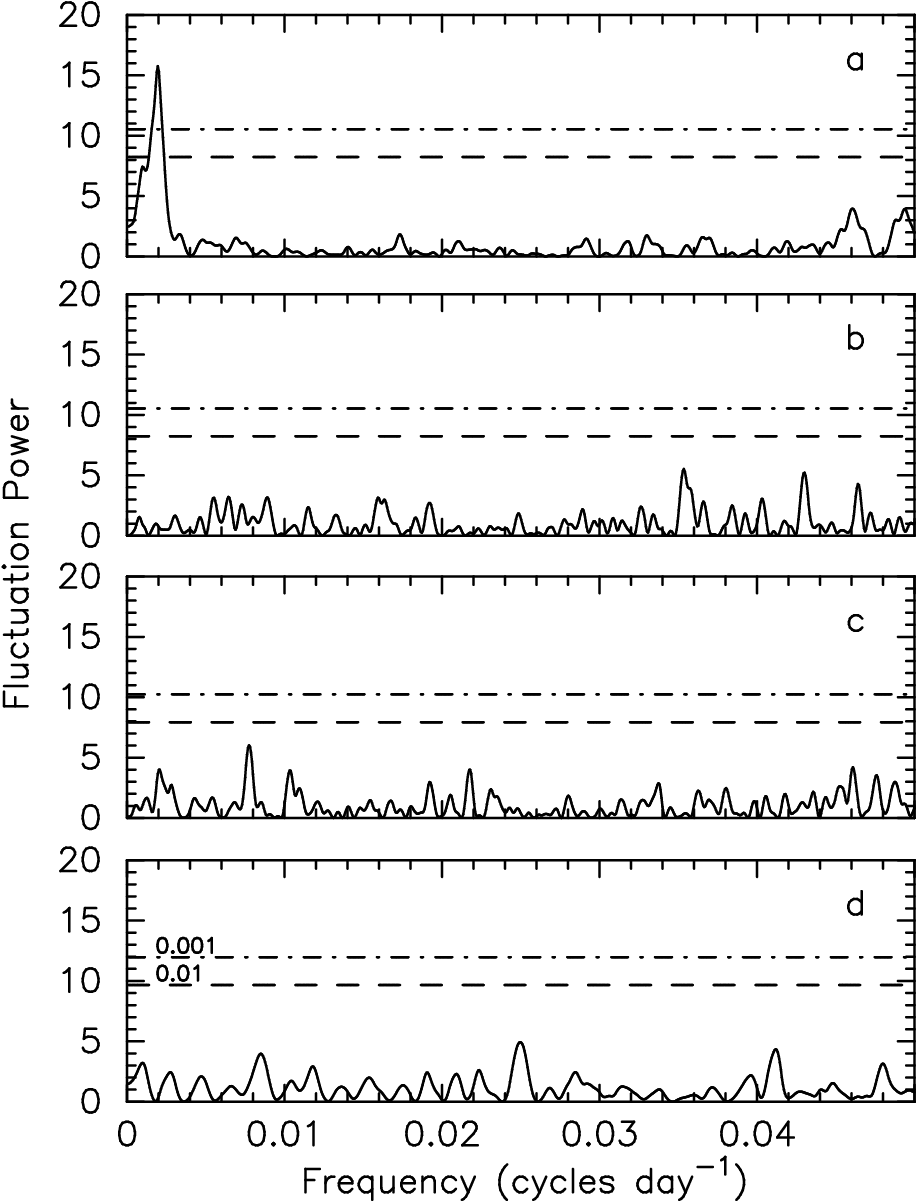}
\caption{The Lomb-Scargle periodograms of (a) Radial velocities (b)  Bisector Velocity Span (c) H$_{\alpha}$ equivalent width and (d) ASAS \citep{1997AcA....47..467P} V photometry of BD +14 4559. The levels of FAP=1.0 $\%$ and 0.1$\%$ are shown.}
\end{figure}

\begin{figure}
\includegraphics[angle=0,scale=0.60]{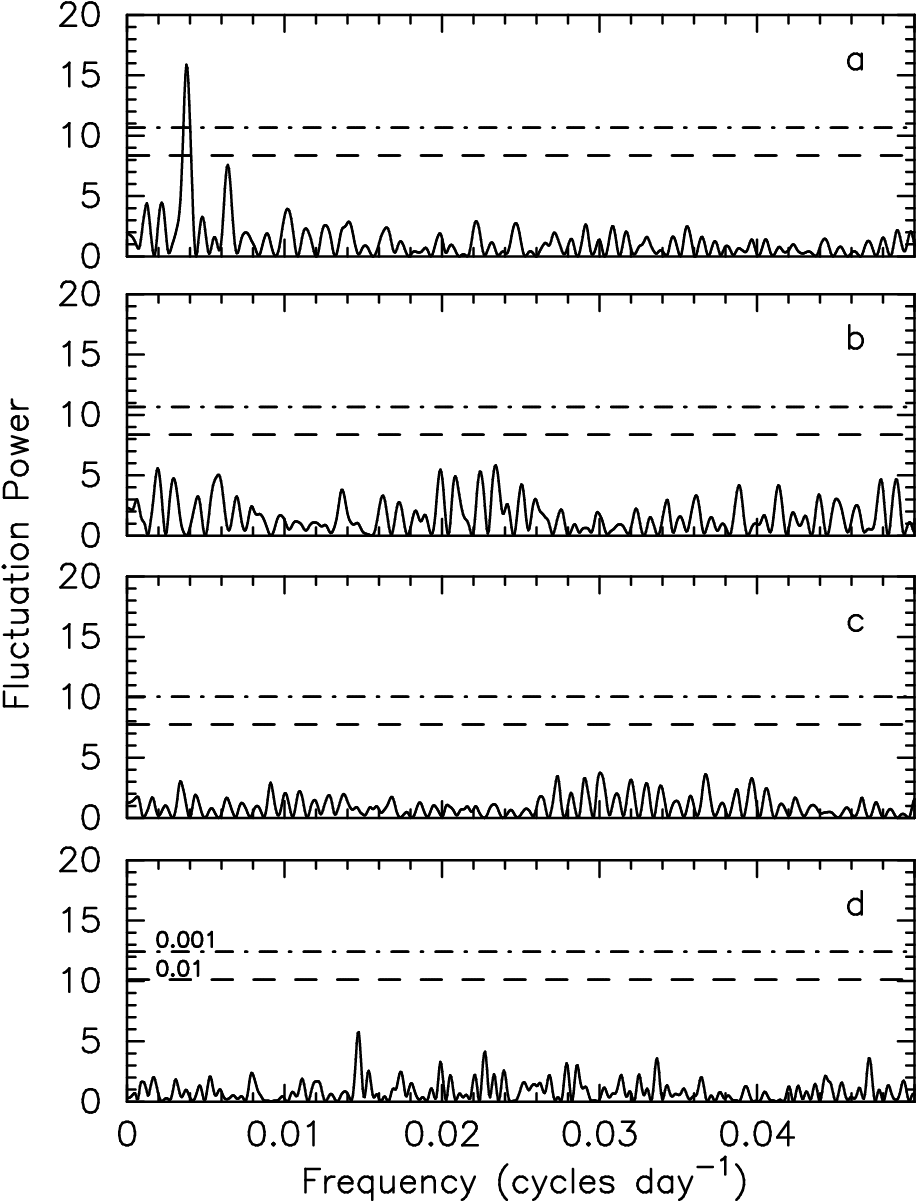}
\caption{The Lomb-Scargle periodograms of (a) Radial velocities (b)  Bisector Velocity Span (c) H$_{\alpha}$ equivalent width and (d) Tycho \citep{1997ESASP1200.....P} V$_{T}$ photometry of HD 240210. The levels of FAP=1.0 $\%$ and 0.1$\%$ are shown.}
\end{figure}

\begin{figure}
\includegraphics[angle=0,scale=0.60]{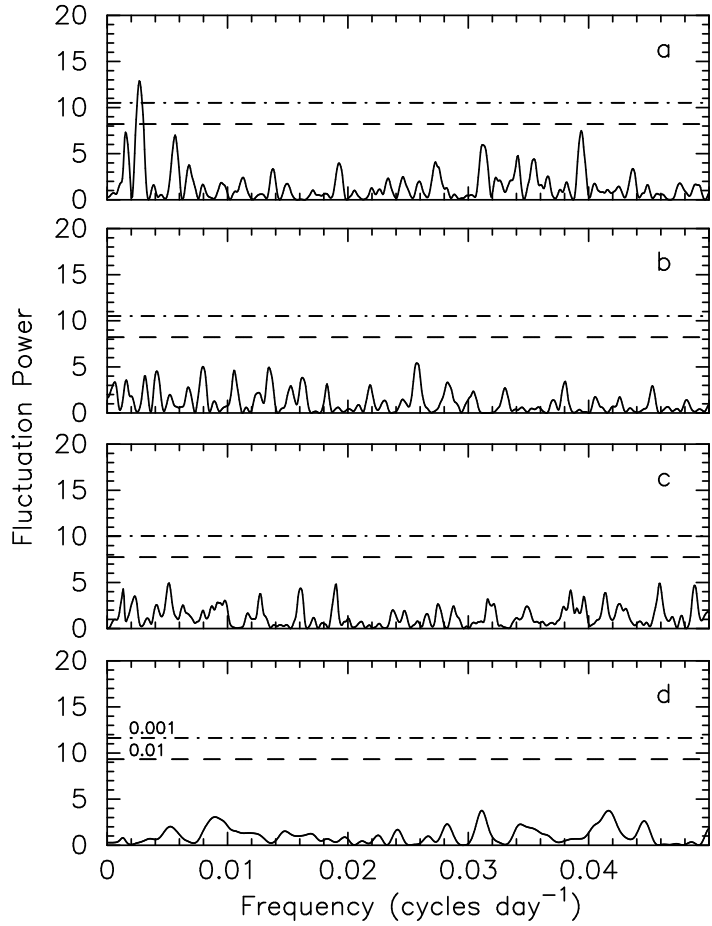}
\caption{The Lomb-Scargle periodograms of (a) Radial velocities (b)  Bisector Velocity Span (c) H$_{\alpha}$ equivalent width and (d) ASAS \citep{1997AcA....47..467P} V  photometry of  BD +20 2457. The levels of FAP=1.0 $\%$ and 0.1$\%$ are shown.}
\end{figure}

\begin{figure}
\includegraphics[angle=-90,scale=0.60]{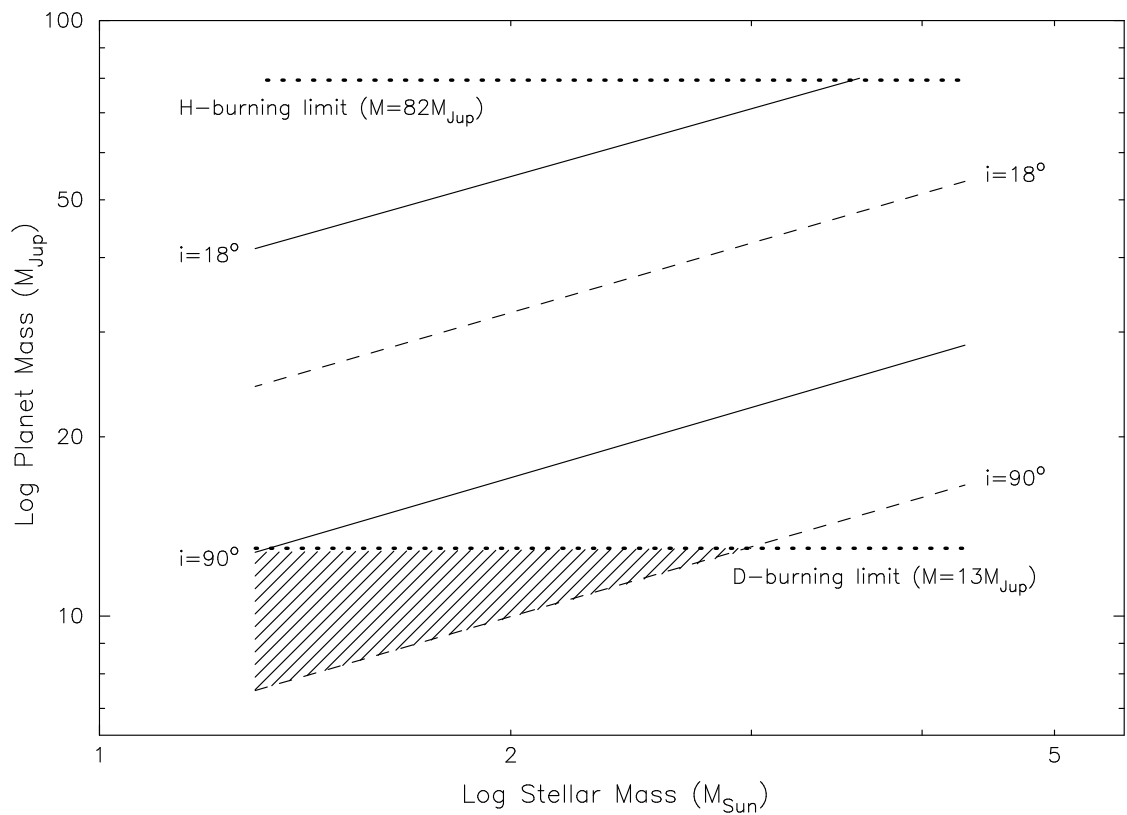}
\caption{Constraints on the masses of substellar companions to BD +20 2457. The solid lines delimit the masses of the inner companion computed from the mass function over the estimated 1.3-4.3 M$_\odot$ range for the stellar mass and orbital inclinations from 18$^\circ$ to 90$^\circ$. The dashed lines have the same meaning for the outer companion. The dotted lines constrain the theoretical mass range for brown dwarfs.}
\end{figure}

\end{document}